# The Global Escape Velocity Profile and Virial Mass Estimate of The Milky Way Galaxy from Gaia Observations

Jeffrey M. La Fortune
1081 N. Lake St. Neenah, WI 54956  forch2@gmail.com
17 October 2022

*Abstract*
Gaia HyperVelocity Star (HVS) kinematic observations favor a local escape velocity of ~700 km/s, nearly forty-percent greater than conventional estimates. Combining HVS and dwarf galaxy satellite data reveal the global escape velocity profile for the Galaxy smoothly traces an unbroken Keplerian decline from the central bar to the most remote satellite galaxy. We reveal a robust upper bound in baryonic mass discrepancy (maximal relative accelerations) linked to the virial theorem and obtain a fundamental and universal mass discrepancy-acceleration relation for virialized compact cosmic objects.

*Introduction*
Understanding and quantifying the Galaxy's escape velocity profile is paramount in determining its mass. In this work, we employ Gaia's precision kinematical data and the virial theorem to construct a new mass model consistent with the underlying thermodynamics. We employ the Mass Discrepancy-Acceleration Relation (MDAR) made popular by McGaugh, Lelli, and Schombert in their seminal paper establishing the Radial Acceleration Relation (herein termed the MLS RAR) derived from the Spitzer Photometry & Accurate Rotation Curves (SPARC) disk galaxy database (Lelli, 2016) (McGaugh, 2016).

MLS revealed a tight 1:1 correlation between baryonic and observed accelerations and near zero dependence with other galactic parameters. Within the MOdified Gravitational Dynamics (MOND) framework, the RAR is a natural outcome as is the Baryonic Tully-Fisher Relation (BTFR) (Milgrom, 1983). In ΛCDM, adherence to these scaling relations is not as distinct and the RAR is often used as a check to ensure consistency between dark matter halos and baryon phenomenology. An argument has been made claiming NFW halo properties also naturally lead to the MLS RAR (Navarro, 2017).

While the MLS RAR is based on acceleration derived from 'cold' ordered motion disk kinematics, our proposed scaling relation is derived from loosely bound or 'hot' Hyper-Velocity Stars (HVS) and the Galaxy's dwarf galaxy satellite population. This approach is an independent measure of galactic mass without resorting to disk components to obtain it. We make extensive use of the MDAR to translate kinematics to relative accelerations and mass discrepancies to draw out the virial scaling relation. We then demonstrate the universal nature of this virial scaling relation by extending it to a sample of galaxy clusters with previously determined RARs.

Many Galactic mass estimates have been conducted with results dependent on tracer selection, spatial location, methodologies, and models (Gallo, 2022). Our approach employs tracers at the mutual interface separating the thermodynamic system from its surroundings (near or at local escape velocities). Our work associates the notion of 'missing mass' with the kinetic energy content of virialized, stationary compact objects and find the virial theorem may offer a solution framework for this mass discrepancy without 'new' physics.

*The Mass Discrepancy/Acceleration-Virial Theorem Link*

The MDAR provides a unique perspective and tool to explore galaxy models and baryonic scaling relations utilizing observed ($a_{Obs}=V_{Obs}^2(r)/r$) and baryonic ($a_{Bar}=GM_{Bar}(<r)/r^2$) acceleration data. We make use of the mass discrepancy parameter $D=(a_{Obs}/a_{Bar})$ and equivalent forms ($V_{Obs}^2/V_{Bar}^2$) and ($M_{Obs}/M_{Bar}$).

The total gravitational potential and virial mass are two properties crucial in the physical description for any compact, gravitationally bound system. For a galaxy in thermodynamic equilibrium, the virial theorem relates its time-averaged total potential (U) to kinetic (T) energies simply as 2T=-U. This theorem also serves as the fundamental basis for many mass estimator models found in the literature (An, 2011). For galaxies, disk properties provide input for the virial theorem and dynamic mass via $M_{Dyn}=V_C^2 R_D G^{-1}$ (circular velocity $V_C$ and radius $R_D$). These same parameters are also used to calculate the Baryonic Tully-Fisher Relation (BTFR), a principal tenet of MOND.[1] Likewise, galaxy disk properties and especially rotation curves have provided a critical constraint in NFW dark matter halo model fits. For the Galaxy, a new halo fitting constraint is introduced based on Gaia kinematics for sample sets not associated with the disk.

Prior to Gaia, only a few extreme velocity stars were identified and those had uncertain or missing proper motions and/or 3D velocities. For many years, black hole 'ejection' remained the leading candidate as calculations demonstrated adequate velocity boost to launch stars beyond Galactic $V_{Esc}$ (Hills, 1988) (Marchetti, 2022). Following this lead, additional 'event' driven ejection mechanisms have been proposed, most requiring very specific (and rare) conditions that have the capacity to launch stars to Galactic escape velocity. As data quality and numbers increased, proposed ejection mechanisms have come to include substructure where close gravitational interactions are more likely. In addition to the central black hole, HVS nurseries are thought to include star clusters, infalling satellites, stellar binaries, accreting dwarf galaxies, and supernovae explosions. With Gaia's precision, HVS orbits can now be back-integrated in time with results indicating a sizeable fraction potentially originating from the disk; the main stellar repository (Marchetti, 2021) (Irrgang, 2018).

In related HVS work, Li also points to galactic substructure as having the necessary conditions to launch stars to hyper-velocities (Li, Q-Z, 2022). Supporting Li's assertion are complementary studies finding very few, if any, emanating from the black hole. In his work, Marchetti discovered a sizable HVS fraction with trajectories having no obvious connection to the Galaxy, deeming these of 'extragalactic' origin. We propose these orphans are fully relaxed stars on highly randomized orbits that no longer reflect initial conditions/point of origin. The bulk of the HVS population may be better described as a time-averaged probability distribution related to open-system thermodynamic processes that support the Galaxy's current quiescent quasi-equilibrium state. The ultimate bulk speeds and orbital properties of the HVS population are highly chaotic making it very difficult to predict initial 'launch' conditions or place of origin with high confidence.

---

[1] Utilizing the 'scaling' model, we have previously introduced a generalized form (gMVR) for the BTFR eliminating $a_0$ from the zero-point expression. Our relation is exactly calculatable from available data as $V_C^4=(\pi G^2 D\Sigma_{Dyn})^{-1}M_{Bar}$ for disk parameters: mass discrepancy (D), and disk dynamic surface mass density ($\Sigma_{Dyn}=M_\odot/\pi R_D^2$) (La Fortune, 2021). Plotting the acceleration ansatz ($\pi G D\Sigma_{Dyn}$) [LT$^{-2}$], In D-$\Sigma_{Dyn}$ space, regression analysis reveals the presence of a common acceleration scale close to $a_0$. Rather than a universal value attributed to MOND, measured accelerations are dependent on individual galactic properties.

*Milky Way Galaxy MDAR Template and 'Scaling' Model*
The Milky Way Galaxy has become a treasure-trove of kinematic information all due to the Gaia mission (Prusti, 2016) (Brown, 2018) (Brown, 2021). In this work we leverage this precision data and analyze the motions of HyperVelocity Stars (HVS) occupying the inner stellar halo and the dwarf satellite galaxy population surrounding the disk (Du, 2019) (Fritz, 2018).

In Figure 1, the MDAR provides a template linking accelerations with associated mass discrepancies based on our treatment of the observed kinematics as a virial phenomenon. The basic virial 'scaling' model consists of two dynamics. The first (green dash) is disk dynamic mass ($M_{Dyn}$) and virial energies 2T=-U. The second (red dash) provides a virial mass ($M_{Vir}$) estimate based on local escape velocity, constant mass discrepancy D=12.1, and T=-U.[2] To simplify the model for our example, we fix dynamic mass to the average ΛCDM cosmic baryon fraction $f_b$=0.17 (D≈1/$f_b$), acknowledging the wide diversity exhibited in the local disk galaxy population. The 'scaling' model fully encloses all dynamic mass ($M_{Dyn}$=0.5x$10^{12}$ $M_\odot$) and baryon mass ($M_{Bar}$=0.085x$10^{12}$ $M_\odot$) within the Galaxy's 80 kpc diameter disk.

Local escape velocities correspond to a virial mass $M_{Vir}$=1.0x$10^{12}$ $M_\odot$ ($M_{Bar}$ x 12.1). This estimate is in close agreement with many past surveys – see Figure 8 (Bird, 2022). In Figure 1 below, we probe the inner halo with Du's HVS sample (blue cross) and outer/extended regions with (high-quality) dwarf satellite galaxy data (blue points) from Fritz. We also present the MLS RAR (gray solid) and our dark matter halo fit based on the simple NFW model (gray dash) and total acceleration with baryon support (black solid).

We make use of Du's HVS selection criteria and sophisticated statistics linking MDAR mass discrepancies with an object's probability of being gravitationally bound to the Galaxy. In Du's HVS analysis, local escape velocity is defined where the probability of HVS being unbound to the Galaxy is fifty-percent ($P_{UB}$≈50%). This escape velocity is consistent for the Galaxy's dynamic mass D=5.9, when it should be fixed to virial mass D=12.1. In revising Du's definition, our local escape velocity matches the global dynamic traced by the dwarf satellite galaxy sample. This revision effectively doubles Galactic mass and the expected increase in escape velocities.[3] In this scenario, nearly all satellite galaxies in our sample can be considered loosely bound, but long-lived residents of the Galactic system. In support, we cite a complementary survey consisting of HVS ($V_{GC}$>450 kms$^{-1}$) obtained by SDSS/APOGEE finding most are gravitationally bound halo stars (Quispe-Huaynasi, 2022).

In Figure 1 we observe most tightly bound HVS align along constant mass discrepancy D=5.9, the Galaxy's dynamic mass. Projecting this to lower accelerations shows Du's escape velocity reference corresponds to the disk's circular velocity at edge (open red circle). In order to maintain a reasonable escape velocity ($V_{Esc}$=√2$V_C$) at this radius, the decline is shallower than Keplerian and closer in form to a NFW dark matter halo.

---

[2] This mass discrepancy value was established several years ago from a 'pre-Gaia' survey of the stellar halo. This value has been used in previous versions of the 'scaling' model and remains an accurate reference point in this age of Gaia (King III, 2015) (La Fortune, 2019). D=12.1 is an arbitrary value but universal for any virialized structure. The effective or observed accelerations may be lower than this expectation due to intrinsic dispersion and the strong upper cut-off.

[3] Deason recently obtained a local escape velocity $V_{Esc}$=528 kms$^{-1}$ which is very close to Williams $V_{Esc}$ =521 kms$^{-1}$ (Deason, 2019) (Williams, 2017). In detail, we find that Deason's total Galactic mass estimate is almost double Williams suggesting this is still an unsettled area of interest.

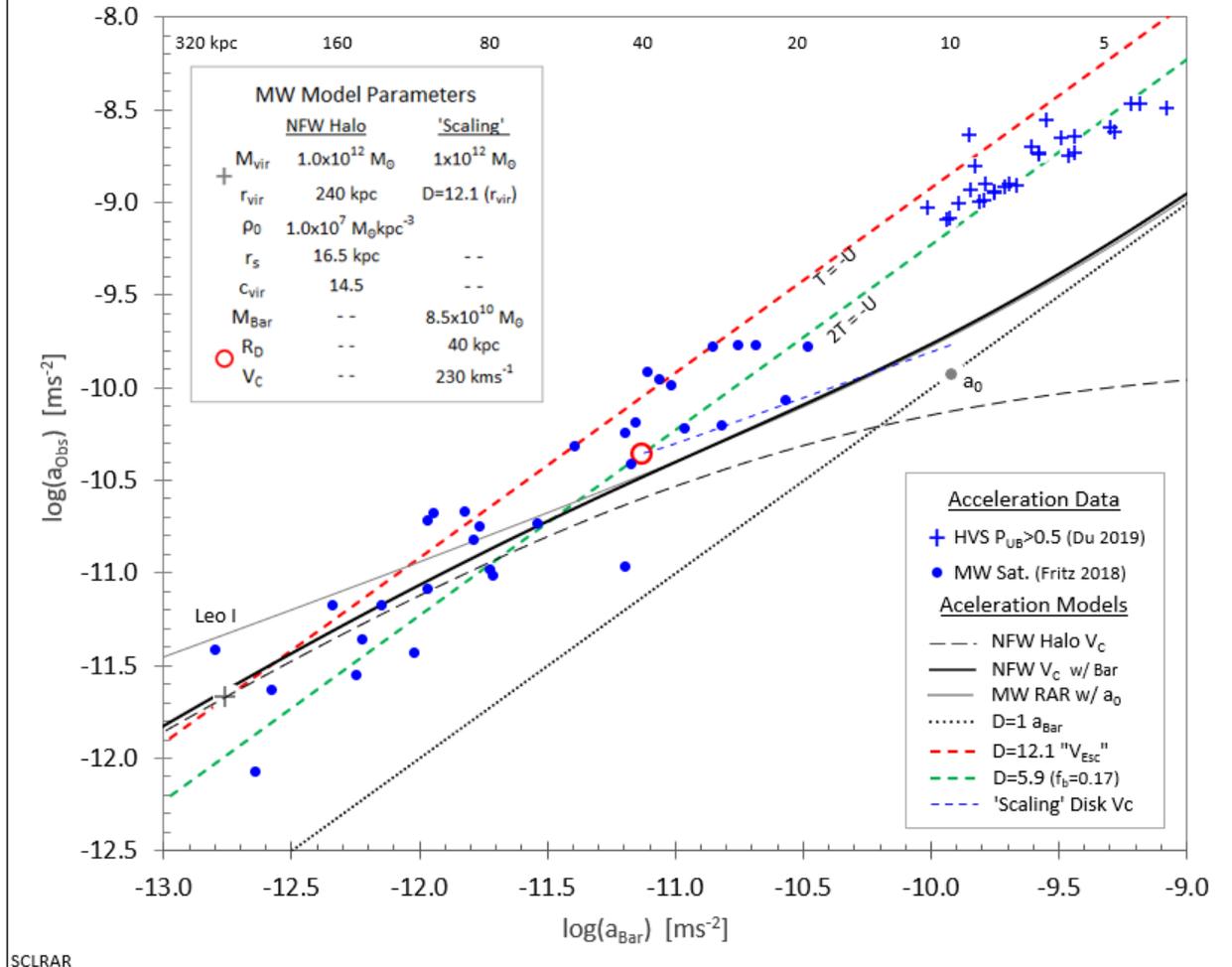

*Figure 1: The MDAR for the Milky Way galaxy with all pertinent halo and 'scaling' model parameters provided in the two keys. Virial energy conditions are represented as constant mass discrepancy D=12.1 (red dash) related to escape velocity (T=-U) and dynamic mass D=5.9 (green dash) satisfying the virial theorem 2T=-U. The Galaxy's disk 'scaling' radius is $R_D$ (red open circle) and constant rotation velocity $V_C$ (short blue dash). A NFW dark matter halo fit is presented with (black solid) and without (gray dash) baryon support. Our halo fit conforms to Navarro's NFW-MDAR relation with peak inner halo acceleration ≈$a_0$ and outer halo acceleration $\log(a_{min}/ms^{-2})$=-11.0 equivalent to dynamic mass proxy D=5.9. If satellite galaxy Leo I is considered a bound object, host halo virial mass increases to 1.6x our estimate. Both NFW halo and the MLS RAR (gray solid) trace the inner-disk velocities well but 'overshoot' the underlying steeper Newtonian dynamic in the low acceleration regime. Sources – (Navarro, 1997) (Boylan-Kolchin, 2013) (McGaugh, 2016) (Navarro, 2017) (Du, 2019) (Fritz, 2018)*

Figure 1 provides an extended view of the Galaxy's acceleration envelope. Most notable is the near absence of HVS and dwarf satellites with mass discrepancies D>12.1, escape velocity's robust upper bound. We find the majority of satellite galaxies are loosely bound near maximal acceleration consistent with our definition of escape velocity.

In the above figure, we fit a simple NFW dark matter halo with baryon velocity (acceleration) support inferred from the HVS sample. We fix virial radius $r_{Vir}$=240 kpc, a nominal value for the Galaxy's size and the halo's escape velocity (acceleration) at $r_{vir}$ to D=12.1 (gray cross).[4] Our NFW scale parameters are reasonable and the rotation curve fit is comparable to previous models.

Our baryon disk also Keplerian and maximal (D=1 <6 kpc) resulting in excessive disk velocities at small radii due to $1/\sqrt{r}$ functionality. We do not account for NFW halo contraction due to the presence of baryons. Even with significant changes in the density gradient, contracted halos still achieve good fits to observational constraints at the expense of diminished baryon content (Cautun, 2020) (Li, P., 2022). Contracted halos are no more challenging to fit than conventional halos, but may result in submaximal disks at odds with SPARC galaxies with circular velocities greater than 200 kms$^{-1}$. Here, we take a different approach fixing baryon support inferred from observation and then performing the halo fit to match the Galactic rotation curve and virial boundary provision described above (Navarro, 1997). Note that we are not advocating for the presence of dark matter, but illustrating the utility of the Gaia observations to guide dynamic models of the Galaxy within the ΛCDM framework.

Turning to the modified gravity model, the MLS RAR (with $a_0$) underestimates the Galaxy's observed circular rotation velocity at the disk's edge. Although easily rectified with a twenty-percent increase in the characteristic acceleration scale, this solution resides outside the MONDian framework. (In MOND expressions involving the BFTR, a geometric normalization factor is introduced that effectively increases the characteristic acceleration scale to better match observations.) The MLS RAR extends beyond the disk with strong long-range velocity support to 'indefinite radii.' This results in the relatively shallow acceleration decline (gray solid) with most satellite galaxies falling below this expectation. A unique property of MOND is that local accelerations are affected by external gravitational fields, termed the External Field Effect (EFE). In the low acceleration regime, the EFE progressively steepens the shallow slope of the MOND RAR to one emulating a Keplerian profile aligning just below global escape velocity D~12, see Figure 10 profiles (Chae, 2020). Perhaps MOND is adding unnecessary complication to an inherently simple dynamic.

*Milky Way Galaxy 'R-V' Space and Escape Velocity Profile*
In this section, we use the same data in Figure 1, but plotted in the Radius-Velocity (R-V) space to gain a spatial perspective of observed velocities and trends. Galactic parameters are stellar disk ($R_P$), HI gas disk ($R_D$), and rotation velocity $V_C$ (blue dash). Galactic dynamic mass and escape velocity is modeled as an unbroken Keplerian decline.[5] In addition to Du and Fritz, three additional sources are included; Li HVS kinematics (red points), Eilers stellar circular rotation velocity (black points) and a virial mass estimate from Bird utilizing non-rotating BHB halo stars (open black square) (Li, Q-Z, 2022).

---

[4] This is one of many kinematically motivated boundary definitions separating Galactic systems from their surroundings. Several definitions related of ΛCDM cosmology have emerged including dark matter halo static mass, splashback, and turnaround, each relying on the condition of vanishing radial velocities (White, 2001) (Cuesta, 2008) (Korkidis, 2020).

[5] Previous versions of the Galaxy 'scaling' model included a pseudo-isothermal region (constant $V_{Esc}$) bridging the inner and outer disk. Gaia exposed an inconsistency in the radial dynamic with introduction of the simplified Newtonian escape velocity profile (Figures 1 and 2 are corrected). While the basic model has shortcomings, it has proven adequate for exploring and describing disk galaxy (and galaxy cluster) acceleration behavior from a global perspective.

We also compare our escape velocity profile to an earlier power law model (gray dot-dash) often used as an exemplar in the literature (Williams, 2017). We include King's total peak stellar halo velocity dispersion value (red cross) obtained at 23 kpc (King III, 2015). In the pre-Gaia era, this served as a $V_{Esc}$ benchmark and basis for D=12.1 in our model. This mass discrepancy value is 'semi-analytic' and does not depend on the physical properties of individual galaxies.

Q-Z Li's HVS selection criteria include those halo stars having $V_{GSR}$>300 kms$^{-1}$ and low metallicities, most of which have not been surveyed in the past. Du's and Li's data occupy the same R-V phase space, lending confidence in our interpretation of the kinematical data. As in the previous figure, we show radial velocity support for our NFW halo fit (gray dash) contrasted to one obtained by Jeans modeling non-rotating stellar halo BHB stars corroborating our result (open black square) (Bird, 2022).

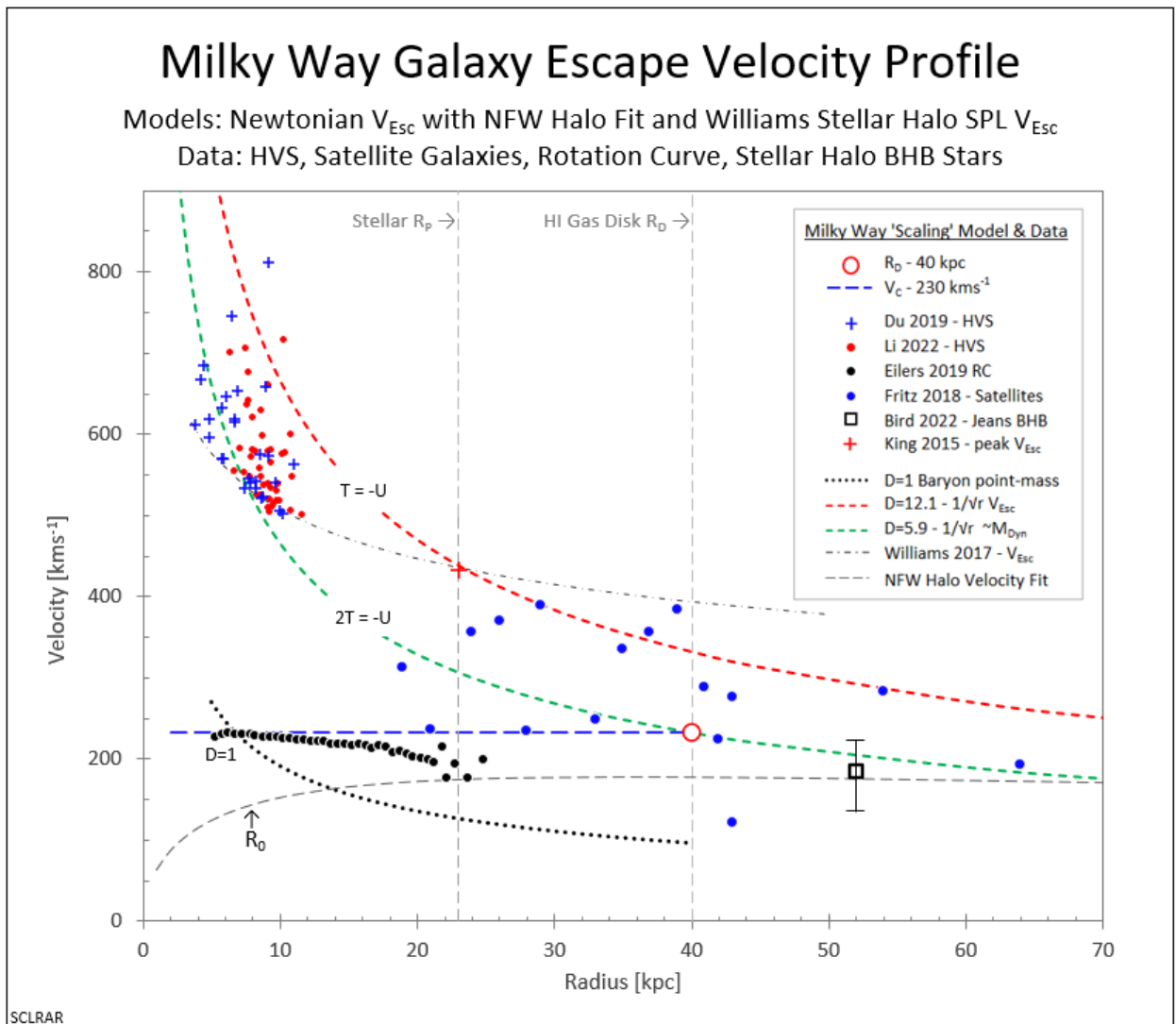

*Figure 2: Milky Way R-V template with kinematic data from five surveys: Du - HVS (blue cross), Li - HVS (red points), Fritz - dwarf satellites (blue points), Eilers - circular disk velocity (black points), and King - peak stellar halo velocity dispersion (red cross). The Galaxy's global escape velocity profile follows a Keplerian decline (red dash). A 'pre-Gaia' $V_{Esc}$ profile estimate (gray dot-dash) obtained by Williams is shown for comparison. Bird's total mass estimate at 52 kpc (open black box/error bars) matches our NFW halo fit (gray dash) with virial mass $M_{Vir}$=1.0x10$^{12}$ $M_\odot$. Sources – (King III, 2015) (Williams, 2017) (Du, 2019) (Eilers, 2019) (Bird, 2022) (Li Q-Z, 2022)*

In Figure 2, both HVS samples, inner satellite galaxy kinematics, and King's data infer an uninterrupted Newtonian $1/\sqrt{r}$ escape velocity profile to 70 kpc. This profile continues unabated to 300 kpc – see Figure 1 (Fritz, 2018) and related Figures 2 and 3 (La Fortune, 2020).

We estimate local ($R_0$) escape velocity of the Galaxy is ~700-750 kms$^{-1}$, much higher than Williams' value of 521 kms$^{-1}$ in agreement with contemporary models.[6] We trace Williams lower $V_{Esc}$ estimate linking it to the Galaxy's dynamic mass rather than virial mass. As a consequence, Williams' total mass estimate is an uncharacteristically low $M_{Tot}=0.3\times10^{12}$ $M_\odot$ <50 kpc, approximately one-half our dynamic mass estimate.

With a finite distributed mass in the Galactic center, the $1/\sqrt{r}$ singularity breakdowns at small radii. A linear inward projection of Du's HVS data indicates peak escape velocity may reach 900 kms$^{-1}$ at the Galactic center rivaling pressure-supported galaxy cluster velocity dispersions. Concerning King's mid-disk result, it is in surprising agreement with Gaia considering the uncertainties in the data at the time. This additional stellar halo datapoint offers additional evidence for an unbroken Keplerian escape velocity decline throughout the disk.

Turning to details related to our NFW halo fit, we plot dark matter halo circular velocity support as a function of radius. Between 70 and 100 kpc, NFW halo and D=5.9 dynamic mass estimates converge to $M_{Dyn}\approx0.5\times10^{12}M_\odot$ per expectation. Beyond the figure, dark matter velocity support continues to $r_{Vir}$ for a final enclosed mass $M_{Vir}=1.0\times10^{12}$ $M_\odot$. As mentioned earlier in Figure 2, we plot Bird's 52 kpc enclosed mass estimate $M_{Tot}=(0.41\pm0.26)\times10^{12}$ $M_\odot$ as velocity demonstrating consistency with our dynamic mass estimate. (This particular analysis was selected as it predicts a NFW halo virial mass equal to our value obtained from Gaia data).

Next, we contrast Bird's approach with a complementary analysis by Eilers using tightly bound disk tracers (Eilers, 2019). Ideally, both approaches should provide very similar Galactic mass estimates but find this is often not the case - tracer location and motion can impact mass estimates. Historical estimates derived from the tightly bound inner disk tend to recover lower total mass values than obtained from loosely bound objects at distance. Our improved sampling sidesteps this issue by integrating several precision surveys into an internally self-consistent global dynamic.

As with Bird, Eilers employed the Jeans equation and modeled the stellar disk as a simple exponential. To construct the circular rotation curve, Eilers considered stars within 0.5 kpc of the disk plane and low vertical velocities to avoid stellar halo contamination. Results gave a local rotation velocity of 229 kms$^{-1}$ and a gently declining profile extending to 20 kpc. Using a baryon model consisting of a Plummer bulge and disk, Eilers estimated the Galaxy's NFW virial mass at $M_{Vir}=7.25\times10^{11}$ $M_\odot$ and halo parameters $r_S=14.8$ kpc, $r_{Vir}=189$ kpc, and $\rho_0=1.1\times10^7$ $M_\odot$kpc$^{-3}$.

As an internal check, we compare Eilers to another viral mass estimate also based on disk kinematics (Ablimit, 2020). Ablimit's analysis corroborates Eilers result with a NFW mass $M_{Vir}=8.22\times10^{11}$ $M_\odot$ and $r_S=17.7$ kpc, $r_{Vir}=192$ kpc, and $\rho_0=1.2\times10^7$ $M_\odot$kpc$^{-3}$. While our NFW halo is slightly larger and heavier than obtained from the Galactic rotation curve, our scale parameters are compatible with these two most recent estimates.

---

[6] Compared to a point-mass, radial disk extension due to angular momentum infers a shallower escape velocity decline exemplified by Williams and others – see Figure 2 (Du, 2019). Contrary to previous models, we recover the $1/\sqrt{r}$ profile with a magnitude consistent with the virial mass of the Galaxy.

Although not shown in Figure 2, the MLS RAR exhibits constant circular velocity extending well past 70 kpc. The downwards deflection in the RAR due to the EFE starts becoming significant beyond 80 kpc with Newtonian alignment achieved at ~200 kpc.

*The Universal Nature of the Virial Theorem*
The virial theorem represents perhaps the most basic structural relationship that any thermodynamic system must obey. As such, disk galaxies and galaxy clusters should demonstrate similar kinematic behavior the system/surroundings boarder (escape velocity). More specifically, cluster MDARs must present the same strong upper mass discrepancy bound (D=12.1) exhibited by the Galaxy.

With recent advancements in cluster X-ray detection/imaging and gravitational lensing techniques coupled with improved hydrostatic models, several 'tests' have been conducted challenging MOND's claim for a universal RAR and a specific characteristic acceleration scale $a_0$. To this end, Chan performed a semi-empirical analysis of the inner cores of clusters and derived an analytic expression for direct comparison to the MLS RAR (Chan, 2022a). While this work conclusively demonstrated great tension between galaxy and cluster RARs, observed accelerations rarely exceeded the aforementioned virial constraint – see Chan Figure 5.

In supporting work, Chan predicted galaxy clusters should demonstrate nearly constant mass discrepancy (D≈8.70±3.42) throughout the bulk of their structure (Chan, 2019). Chan attributed this behavior to unspecified 'interplay' between baryons and dark matter (our kinetic energy ansatz). Irregardless of motivation, it very accurately reproduces observations with a range between D=6.3 to 11.5 found to be in agreement with several independent observations presented below.

In Figure 3 we combine galaxy cluster RAR regression fits (colored solids with key) from several recent surveys (Chan, 2019) (Chan, 2020) (Tian, 2020) (Pradyumna, 2021a) (Pradyumna, 2021b) (Eckert, 2022). We are interested in general cluster magnitudes and trends with comparison to regressions obtained from the Galaxy's HVS (light blue points) and satellites galaxy (light blue dash) samples. The 'virial framework' is the same illustrated in Figure 1 with 2T=-U dynamic mass (green dash) and T=-U virial mass (red dash) acceleration profiles.

The MLS RAR (gray solid) and $a_0$ (gray point) provide a reference and highlight the discrepancy between galactic and cluster RARs. The MLS RAR is derived from the galactic disk, its most massive, luminous component with its most prominent kinematic feature, the circular rotation curve. Likewise, galaxy clusters RARs are also obtained from their most luminous component but in this case it is highly thermalized gas. Whereas disk rotation is an accepted parameter to model acceleration behavior of disk galaxies, pressure-supported clusters require an alternative kinematic measure – velocity dispersion of the bulk gas component.

MOND's assertion that modified gravitational law is universal and present in both structures can be inferred by the very close relationship demonstrated between the BTF and the Baryonic Jackson-Faber (BJF) relations. We contend that the BTFR and BFJR are global relations satisfying the virial theorem as evidenced by a universal and predominate escape velocity constraint – see Appendix C (La Fortune, 2021). A simple comparison of the RARs may discount MOND, but the galactic MLS RAR offers a complementary and powerful scaling relation not present in galaxy clusters.

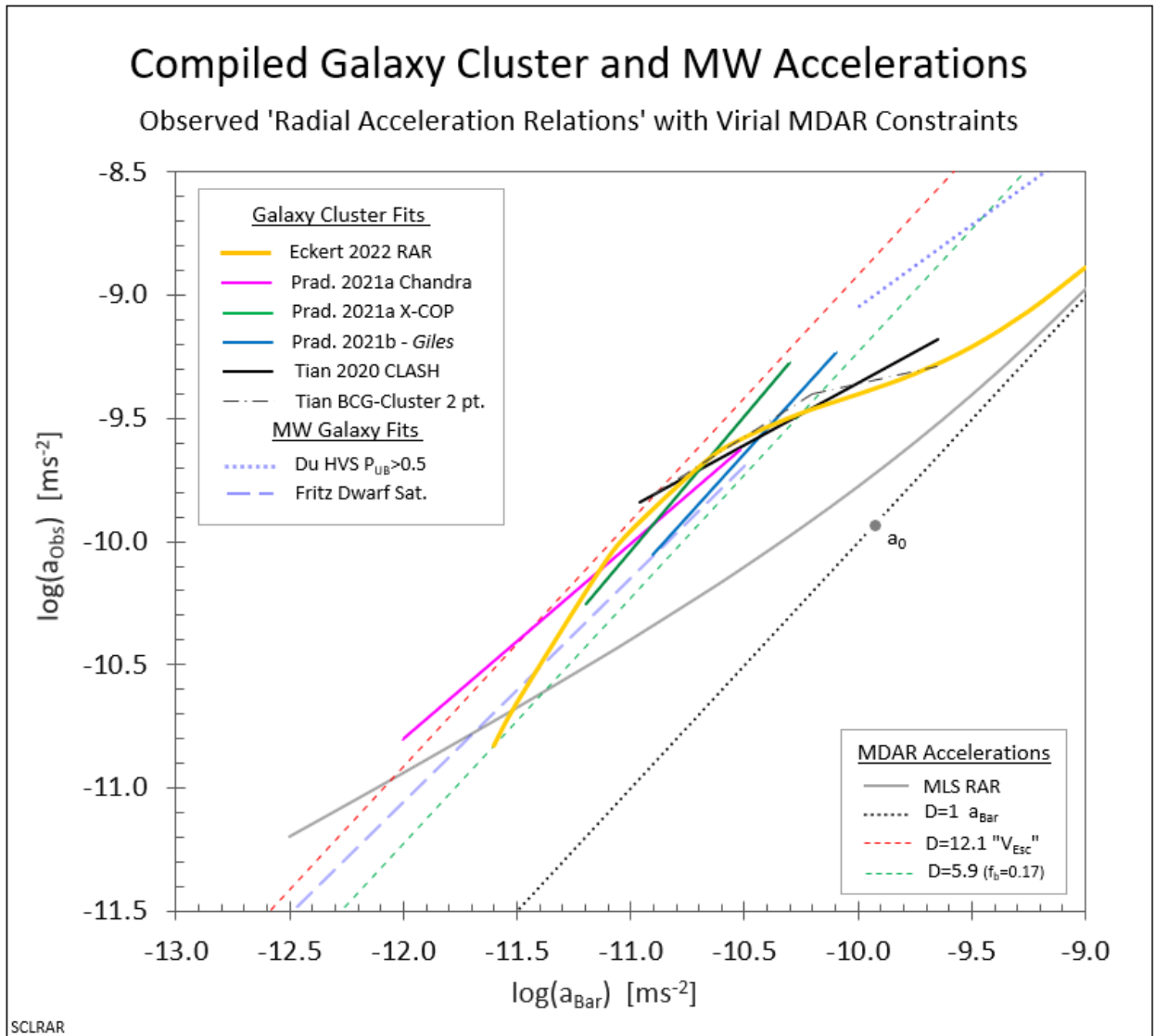

*Figure 3: MDAR plot of compiled galaxy cluster radial acceleration results with comparison to Galaxy HVS and satellite galaxies – each identified in key. Note the similarities in the acceleration footprint and the robust acceleration/escape velocity 'cut-off' at D=12.1. The MLS RAR (gray solid) underestimates cluster accelerations over a wide range due to tightly bound stellar disks (D<5.9) compared to loosely bound cluster gas (5.9<D<12.1). The combined data infer a common mechanism regulating and moderating galactic and cluster acceleration behavior – thermodynamic equilibrium. Sources - in text and key.*

As with the Galaxy, all cluster RARs are subject to the escape velocity constraint D=12.1. Unlike disk galaxies, clusters are pressure-supported systems with accelerations due to high gas velocity dispersion. This gas is maximally accelerated with most cluster RARs just below D=12.1 over a very wide span of accelerations similar to the Galaxy's loosely bound HVS and dwarf satellite galaxy populations.

In Figure 3, while most cluster RARs align along escape velocity (log) slope γ≈1 close to Chan's analytically derived constant D=8.7, Tian's CLASH RAR (black solid) demonstrates a slope γ≈0.5 consistent with the MLS RAR but having a characteristic acceleration scale nearly 20x $a_0$. In this study, CLASH cluster accelerations were measured at five specific radii (14, 100, 200, 400, and 600 kpc) with the 14 kpc assigned to the Brightest Cluster Galaxy (BCG) and the rest attributed to gas kinematics.

We can gain more information from Tian's data by isolating the BCG from the cluster measured at the larger radii. Breaking these apart and replotting, the two-part acceleration curve (black dot-dash) now closely matches Eckert's more complex cluster RAR (yellow solid) smoothly melding BCG stellar and thermalized gas accelerations (Bilek, 2017) (Chan, 2022b). In Figure 3, the Eckert RAR reverts to D~5.9 (the cosmic baryon fraction) at the observational cut-off in broad agreement with Tian and other surveys.

*Summary*

We argue that thermodynamic processes are responsible for the Galaxy's preferred global mass-energy configuration comprising a disk of constant circular velocity and surprisingly, a Keplerian global escape velocity profile. Without introducing new physics, we propose a new and universal mass discrepancy-acceleration relation linking galactic and cluster kinematic properties to the virial theorem. The concept of "missing mass" is a manifestation of the kinetic energy content inherent in massive compact 'systems' inhabiting the local universe.

*Acknowledgements*

The author greatly appreciates the helpful correspondence and 'raw' data sharing throughout the process of completing this article. We thank authors, arXiv, and professional journals for their willingness to publish research making it available for all that have an interest.

*Bibliography*

Ablimit, I. (2020, May 20). The Rotation Curve, Mass Distribution, and Dark Matter Content of the Milky Way from Classical Cepheids. *ApJ, 895:L12,* https://iopscience.iop.org/article/10.3847/2041-8213/ab8d45/meta.

An, J.H. (2011, May 11). Modified virial formulae and the theory of mass estimators. *MNRAS, 413,1744-1752,* https://academic.oup.com/mnras/article/413/3/1744/964719.

Bilek, M. (2017, Nov 23). Investigating the radial acceleration relation in early-type galaxies using the Jeans analysis. *arXiv, 1711.06335,* https://arxiv.org/abs/1711.06335.

Bird, S.A. (2022, Jul 22). Milky Way Mass with K Giants and BHB Stars Using LAMOST, SDSS/SEGUE, and Gaia: 3D Spherical Jeans Equation and Tracer Mass Estimator. *MNRAS, stac3026,* https://academic.oup.com/mnras/advance-article-abstract/doi/10.1093/mnras/stac2036/6649346?redirectedFrom=fulltext.

Boylan-Kolchin, M. (2013, Apr 24). THE SPACE MOTION OF LEO I: THE MASS OF THE MILKY WAY'S DARK MATTER HALO. *AJ, 768:140,* https://iopscience.iop.org/article/10.1088/0004-637X/768/2/140.

Brown, A.G.A. (2018, Aug 18). Gaia Data Release 2. *A&A, 616, A1,* https://www.aanda.org/articles/aa/abs/2018/08/aa33051-18/aa33051-18.html.

Brown, A.G.A. (2021, Apr 28). Gaia Early Data Release 3. *A&A, 649, A1,* https://www.aanda.org/articles/aa/abs/2021/05/aa39657-20/aa39657-20.html.

Cautun, M. (2020, Apr 17). The milky way total mass profile as inferred from Gaia DR2. *MNRAS, 494, 3, 4291–4313,* https://academic.oup.com/mnras/article/494/3/4291/5821286.

Chae, K-H (2020, Nov 20). Testing the Strong Equivalence Principle: Detection of the External Field Effect in Rotationally Supported Galaxies. *ApJ, 904, 1,* https://iopscience.iop.org/article/10.3847/1538-4357/ac1bba/meta.

Chan, M.H. (2019, Mar 5). *A universal constant for dark matter-baryon interplay.* Retrieved Aug 30, 2020, from Nature - Scientific Reports 9:3570, https://www.nature.com/articles/s41598-019-39717-x

Chan, M.H. (2020, Jan 28). The radial acceleration relation in galaxy clusters. *MNRAS, 492, 4,* https://academic.oup.com/mnras/article-abstract/492/4/5865/5716677.

Chan, M.H. (2022b, May 16). There is no universal acceleration scale in galaxies. *arXiv, 2205.07515,* https://arxiv.org/abs/2205.07515.

Chan, M.H. (2022a, Apr 12). Analytic radial acceleration relation for galaxy clusters. *Phys. Rev. D, 105, 083003,* https://journals.aps.org/prd/abstract/10.1103/PhysRevD.105.083003.

Cuesta, A.J. (2008, Aug 18). The virialized mass of dark matter haloes. *MNRAS, 389, 1, pg 385-397,* https://academic.oup.com/mnras/article/389/1/385/995018.

Deason, A.J. (2019, Mar 2). The local high-velocity tail and the Galactic escape speed. *MNRAS, 485, pg 3515-3525,* https://academic.oup.com/mnras/article/485/3/3514/5368371.

Du, C. (2019, Aug 22). New Nearby Hypervelocity Stars and Their Spatial Distribution from Gaia DR2. *APJ Supplement Series, Vol. 244, Issue 1,* https://iopscience.iop.org/article/10.3847/1538-4365/ab328c/meta.

Eckert, D. (2022, May 2). The gravitational field of X-COP galaxy clusters. *arXiv 2205.01110,* https://arxiv.org/abs/2205.01110.

Eilers, A-C (2019, Jan 25). The Circular Velocity Curve of the Milky Way from 5 to 25 kpc. *ApJ, 871, 120,* https://iopscience.iop.org/article/10.3847/1538-4357/aaf648/meta.

Fritz, T.K. (2018, Nov 13). Gaia DR2 proper motions of dwarf galaxies within 420 kpc: Orbits, Milky Way mass, tidal influences, planar alignments, and group infall. *A&A*, 619, A103, https://www.aanda.org/articles/aa/abs/2018/11/aa33343-18/aa33343-18.html.

Gallo, A. (2022, Jul 22). Probing the shape of the Milky Way dark matter halo with hypervelocity stars: A new method. *A&A, 663, A72,* https://www.aanda.org/articles/aa/full_html/2022/07/aa42679-21/aa42679-21.html.

Hills, J.G. (1988, Feb 25). Hyper-velocity and tidal stars from binaries disrupted by a massive Galactic black hole. *Nature. 331, p687–689*, https://www.nature.com/articles/331687a0.

Irrgang, A. (2018, Jul 17). Hypervelocity stars in the Gaia era - Runaway B stars beyond the velocity limit of classical ejection mechanisms. *Astronomy & Astrophysics, A48,* https://www.aanda.org/articles/aa/full_html/2018/12/aa33874-18/aa33874-18.html.

King III, C. (2015, Nov 10). STELLAR VELOCITY DISPERION AND ANISOTROPY OF THE MILKY WAY INNER HALO. *The Astrophysical Journal, 813:89,* http://iopscience.iop.org/article/10.1088/0004-637X/813/2/89/pdf.

Korkidis, G. (2020, Jul 20). Turnaround radius of galaxy clusters in N-body simulations. *A&A, 639, A122* https://www.aanda.org/articles/aa/abs/2020/07/aa37337-19/aa37337-19.html.

La Fortune, J.M. (2019, Dec 16). The Phenomenological Scaling Relations of Disk Galaxies. *arXiv, 1912.07335,* https://arxiv.org/abs/1912.07335.

La Fortune, J.M. (2020, Aug 25). A Physical Interpretation of Milky Way Galaxy Dynamics from Precision Astrometrics. *arXiv, 2008.10445* https://arxiv.org/abs/2008.10445.

La Fortune, J.M. (2021, Oct 6). On a Generalized Mass-Velocity Relation for Disk Galaxies and Galaxy Clusters. *arXiv, 2110.02684*, p. https://arxiv.org/abs/2110.02684.

Lelli, F. (2016, Nov 11). SPARC: MASS MODELS FOR 175 DISK GALAXIES WITH SPITZER PHOTOMETRY AND ACCURATE ROTATION CURVES. *The Astronomical Journal, 152, 6,* https://iopscience.iop.org/article/10.3847/0004-6256/152/6/157.

Li, P. (2022, Aug 3). Incorporating baryon-driven contraction of dark matter halos in rotation curve fits. *A&A forthcoming*, https://www.aanda.org/component/article?access=doi&doi=10.1051/0004-6361/202243916.

Li, Q-Z (2022, Jul 10). On the origins of Hypervelocity stars as revealed by large-scale Galactic surveys. *arXiv, 2207.04406,* https://arxiv.org/abs/2207.04406.

Marchetti, T. (2021, Mar 1). Gaia EDR3 in 6D: searching for unbound stars in the galaxy. *MNRAS, 503, 1,* https://academic.oup.com/mnras/article-abstract/503/1/1374/6155058.

Marchetti, T. (2022, Jun 29). Gaia DR3 in 6D: the search for fast hypervelocity stars and constraints on the galactic centre environment. *MNRAS, 515, 1, pg 767-774,* https://academic.oup.com/mnras/article-abstract/515/1/767/6620843.

McGaugh, S. S. (2016, Nov 9). The Radial Acceleration Relation in Rotationally Supported Galaxies. *Phys. Rev. Lett. 117, 201101*, https://journals.aps.org/prl/abstract/10.1103/PhysRevLett.117.201101.

Milgrom, M. (1983, Jul 15). A modification of the Newtonian dynamics as a possible alternative to the hidden mass hypothesis. *The Astrophysical Jounal*, 270, 365-270, http://adsabs.harvard.edu/full/1983ApJ...270..365M7.

Navarro, J.F. (2017, Jul 8). The origin of the mass discrepancy–acceleration relation in ΛCDM. *MNRAS, 471, 1841-1848,* https://academic.oup.com/mnras/article/471/2/1841/3939742.

Navarro, J.F. (1997, Dec 1). A Universal Density Profile from Hierarchical Clustering. *ApJ, 490, 2, 493-508,* https://iopscience.iop.org/article/10.1086/304888/meta.

Pradyumna, S. (2021a, Jan -). Yet another test of Radial Acceleration Relation for galaxy clusters. *Physics of the Dark Universe, Vol. 31, 100765,* https://www.sciencedirect.com/science/article/abs/pii/S2212686420304787?via%3Dihub.

Pradyumna, S. (2021b, Sep -). A test of Radial Acceleration Relation for the Giles et al Chandra cluster sample. Physics of the Dark Universe, 33, 100854, https://www.sciencedirect.com/science/article/abs/pii/S2212686421000844

Prusti, T. (2016, Nov 24). The Gaia mission. *A&A, 595, A1*, https://www.aanda.org/articles/aa/abs/2016/11/aa29272-16/aa29272-16.html.

Quispe-Huaynasi, F. (2022, Sep 8). High Velocity Stars in SDSS/APOGEE DR17 . *arXiv, 2209.03560 (submitted to AJ),* https://arxiv.org/abs/2209.03560.

Tian, Y. (2020, Jun 10). The Radial Acceleration Relation in CLASH Galaxy Clusters. *ApJ, Vol 6, No 1*, https://iopscience.iop.org/article/10.3847/1538-4357/ab8e3d/meta.

White, M. (2001, Feb 15). The mass of a halo. *A&A, 367, 1, pg 27-32*, https://www.aanda.org/articles/aa/abs/2001/07/aah2421/aah2421.html.

Williams, A.A. (2017, Mar 2). On the Run: mapping the escape speed across the Galaxy with SDSS. *MNRAS, 468, 2, Pg 2359–2371*, https://academic.oup.com/mnras/article/468/2/2359/3059977.